\documentclass{article}
\usepackage[T1]{fontenc} 
\usepackage[utf8]{inputenc} 
\usepackage{ismir,amsmath,cite,url}
\usepackage{graphicx}
\usepackage{color}

\usepackage{lineno}

\usepackage{amsfonts}       
\usepackage[acronym, hyperfirst=false]{glossaries}

\usepackage{booktabs}       
\usepackage{tikz}
\usepackage{graphicx}
\usepackage[export]{adjustbox}

\usetikzlibrary{shapes, arrows.meta, positioning, shapes.geometric, calc}

\title{Generating Sample-Based Musical Instruments Using Neural Audio Codec Language Models}

\multauthor
{Shahan Nercessian$^{*}$ \hspace{0.5cm} Johannes Imort$^{*}$ \hspace{0.5cm} Ninon Devis \hspace{0.5cm} Frederik Blang} { Native Instruments \\ {\tt firstname.lastname@native-instruments.com} \hspace{0.5cm} \\ {\small $^{*}$Equal contribution}
}

\def\authorname{S. Nercessian, J. Imort, N. Devis, and F. Blang}

\definecolor{color1}{HTML}{320a5e}
\definecolor{color2}{HTML}{781c6d}
\definecolor{color3}{HTML}{bc3754}
\definecolor{color4}{HTML}{fbb61a}
\definecolor{color5}{HTML}{fcffa4}

\begin{document}

\newacronym{ddsp}{DDSP}{Differentiable Digital Signal Processing}
\newacronym{tc}{TC}{timbral consistency}
\newacronym{dac}{DAC}{Descript Audio Codec}
\newacronym{rvq}{RVQ}{Residual Vector Quantization}
\newacronym{midi}{MIDI}{Musical Instrument Digital Interface}
\newacronym{clap}{CLAP}{Contrastive Language-Audio Pretraining}
\newacronym{fad}{FAD}{Fréchet audio distance}
\newacronym{mos}{MOS}{Mean Opinion Scores}
\newacronym{lut}{LUT}{lookup table}
\newacronym{mushra}{MUSHRA}{MUltiple Stimuli with Hidden Reference and Anchor}
\newacronym{t2i}{T2I}{text-to-instrument}
\newacronym{s2i}{S2I}{sample-to-instrument}
\newacronym{ar}{AR}{autoregressive}
\maketitle
\begin{abstract}
In this paper, we propose and investigate the use of neural audio codec language models for the automatic generation of sample-based musical instruments based on text or reference audio prompts. Our approach extends a generative audio framework to condition on pitch across an 88-key spectrum, velocity, and a combined text/audio embedding. We identify maintaining timbral consistency within the generated instruments as a major challenge. To tackle this issue, we introduce three distinct conditioning schemes. We analyze our methods through objective metrics and human listening tests, demonstrating that our approach can produce compelling musical instruments. Specifically, we introduce a new objective metric to evaluate the timbral consistency of the generated instruments and adapt the average Contrastive Language-Audio Pretraining (CLAP) score for the text-to-instrument case, noting that its naive application is unsuitable for assessing this task. Our findings reveal a complex interplay between timbral consistency, the quality of generated samples, and their correspondence to the input prompt.
\end{abstract}

\begin{figure*}[t] 
\vspace{-1em}
\centering
\resizebox{0.95\textwidth}{!}{%
  \begin{tikzpicture}[
    encoderblock/.style={scale=1, shape=trapezium, trapezium angle=75, draw, shape border rotate=270, trapezium stretches=true, minimum width=20mm, minimum height=30mm, draw = color1, fill = color1!10, rounded corners= 3pt, line width = 1pt},
    decoderblock/.style={scale=1, shape=trapezium, trapezium angle=75, draw, shape border rotate=90, trapezium stretches=true, minimum width=20mm, minimum height=30mm, draw = color1, fill = color1!10, rounded corners= 3pt, line width = 1pt},
    block/.style={scale=1, shape=rectangle, draw, minimum width=15mm, minimum height=10mm, rounded corners= 3pt, line width = 1pt, draw = color3, fill = color3!10},
    linear/.style={scale=1, shape=rectangle, draw, minimum width=10mm, minimum height=5mm, rounded corners= 3pt, line width = 1pt, draw = color4, fill = color4!10},
    dot/.style={circle, fill,inner sep=1pt},
    loss/.style={scale=1, shape=rectangle, draw, minimum width=15mm, rounded corners= 3pt, line width = 1pt, draw = color2, fill = color2!10},
    font=\sffamily\small,
    text=black!80,
    ]
    \def\dx{5mm}
    \def\dy{10mm}
 
    \node (base) at (0mm, 0mm) {};

    \node (dac_encoder) [encoderblock, left=0mm of base, dotted] {DAC encoder};

    \node (clap_encoder) [encoderblock, below=1.5*\dy of dac_encoder, dotted, align=center, opacity=0.75, text opacity=1.0] {CLAP audio head\\$E_\mathrm{a}$};

    \node (clap_text_encoder) [encoderblock, below=-1mm of clap_encoder, dotted, align=center, opacity=0.75, text opacity=1.0] {CLAP text head\\$E_\mathrm{t}$};

    \node (clap_text_encoder_input) [left=0.5cm of clap_text_encoder, align=center] {text prompt \\ (inference)};

    \draw [-stealth] (clap_text_encoder_input) -- (clap_text_encoder) node[midway, above] {$\mathbf{t}_k$};


    \node [rotate=90, left=3cm of dac_encoder, anchor=center] (waveform_input_dac) {\includegraphics[width=3cm, height=1cm]{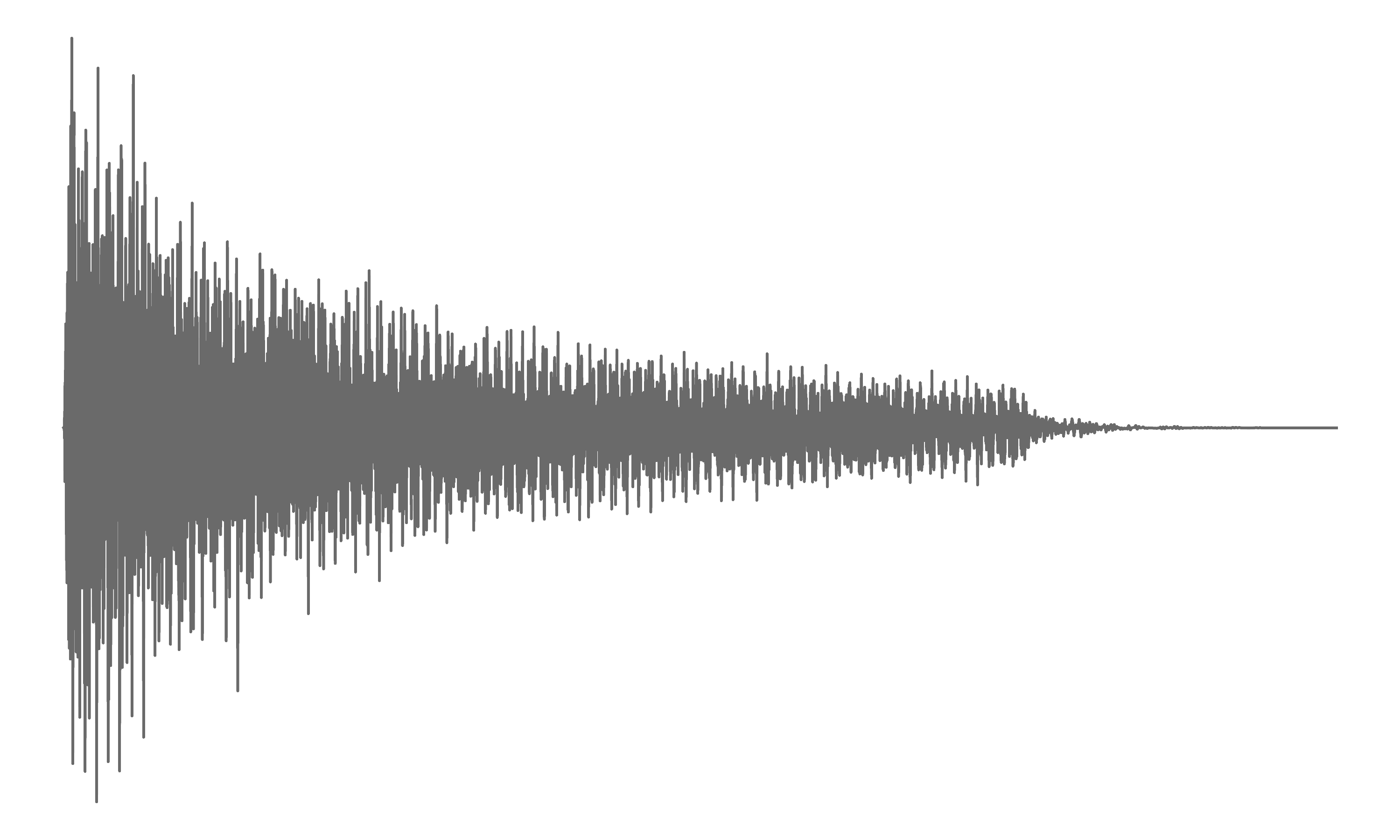}};
    
    \node [right=2.4cm of waveform_input_dac, anchor=0, yshift=-0.5cm] (piano1) {\includegraphics[width=1cm, height=1cm]{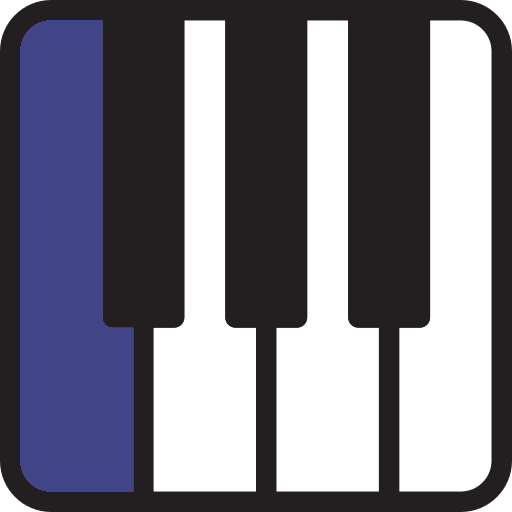}};
    
    \node[rotate=90, anchor=north, align=center] at ($(waveform_input_dac.north) + (-5mm, 0)$) (waveform_input_dac_text) {input waveform \\ (training)};

    \node [rotate=90, left=3cm of clap_encoder, anchor=center] (waveform_input_clap) {\includegraphics[width=3cm, height=1cm]{waveform_grey.png}};
    
    \node [right=2.4cm of waveform_input_clap, anchor=0, yshift=-0.5cm] (piano1) {\includegraphics[width=1cm, height=1cm]{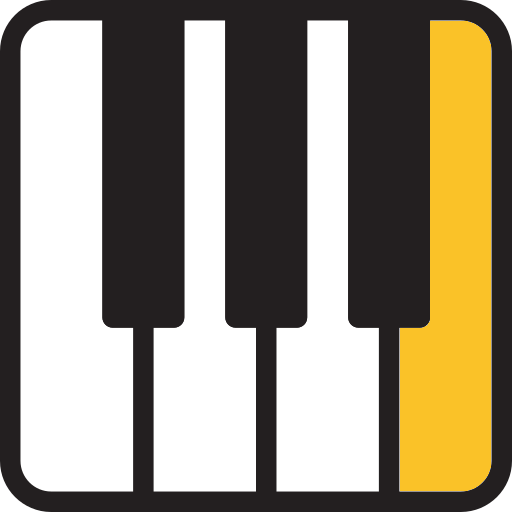}};

    \node[rotate=90, anchor=north, align=center] at ($(waveform_input_clap.north) + (-5mm, 0)$) (waveform_input_dac_text) {audio prompt \\ (training/inference)};


    \draw [-stealth, dashed] (waveform_input_dac.south) --  (dac_encoder) node[midway, above] {$\mathbf{x}_{k}(p, v)$};

    \draw [-stealth] (waveform_input_clap.south) --  (clap_encoder) node[midway, above] {$\mathbf{x}_k(\rho, \nu)$};

    \coordinate (midpoint_clap_enc) at ($(clap_encoder.east)!0.5!(clap_text_encoder.east)$);

    \node[dot, draw=black, fill=none] (endpoint_clap_audio) at ($(clap_encoder.east) + (1.25cm, 0cm)$) {};
    \node[dot, draw=black, fill=none] (endpoint_clap_text) at ($(clap_text_encoder.east) + (1.25cm, 0cm)$) {};

    \coordinate (midpoint_clap_switch) at ($(endpoint_clap_audio)!0.5!(endpoint_clap_text)$);

    \node[dot] (switch_merge) at ($(midpoint_clap_switch) + (1cm, 0cm)$) {};

    \draw [solid] ($(endpoint_clap_audio) - (0.05cm, 0.05cm)$) -- (switch_merge);

    \coordinate (midpoint_switch_arrow) at ($(endpoint_clap_audio)!0.5!(switch_merge)$);
    \draw [->] ($(midpoint_switch_arrow) + (0.25cm, 0.25cm)$) to [bend right=45] ($(midpoint_switch_arrow) - (0.25cm, 0.25cm)$);
    
    \draw (clap_encoder) -- (endpoint_clap_audio) node[midway, above] {$\mathbf{z}_\mathrm{\glsentrytext{clap}, a}$};
    \draw (clap_text_encoder) -- (endpoint_clap_text) node[midway, above] {$\mathbf{z}_\mathrm{\glsentrytext{clap}, t}$};

    \node (rvq_clap) [block, right=2.0*\dx of switch_merge] {\glsentrytext{rvq}};

    \draw [-stealth] (switch_merge) -- (rvq_clap) node[midway, above] {$\mathbf{z}_\mathrm{\glsentrytext{clap}}$};


    \node (rvq_linear) [linear, right=10mm of rvq_clap]{Linear};

    \node (rvq) [block, dotted] at (dac_encoder -| rvq_clap) {\glsentrytext{rvq}};
    
         
    \node (lm) [block, right=3*\dx of rvq, align=center] {Transformer decoder};
    \node (dac_decoder) [decoderblock, right=3*\dx of lm, dotted] {DAC decoder};

    \coordinate (outpoint) at ($(dac_decoder.east) + (1cm, 0cm)$);
        
    \node (waveform_output_dac) [rotate=90, right=2cm of outpoint, anchor=center] {\includegraphics[width=3cm, height=1cm]{waveform_grey.png}};
    

    \def\initialShiftX{2cm} 
    \def\initialShiftY{0cm} 
    \def\shiftStepX{0.2cm} 
    \def\shiftStepY{0.2cm} 
    \def\numWaveforms{5} 
    
    \pgfmathsetmacro{\totalShiftX}{0}
    \pgfmathsetmacro{\totalShiftY}{0}
    \pgfmathsetmacro{\opa}{1.0}
    
    \foreach \n in {1,...,\numWaveforms}{
        \pgfmathsetmacro{\totalShiftX}{\initialShiftX + \shiftStepX*(\n - 1)}
        \pgfmathsetmacro{\totalShiftY}{\initialShiftY + \shiftStepY*(\n - 1)}
        \pgfmathsetmacro{\opa}{\opa*(0.5^(\n - 1))}
        
        \node at ([xshift=\totalShiftX,yshift=\totalShiftY]outpoint) [rotate=90, anchor=center] {
            \begin{tikzpicture}
                \node [opacity=\opa] {\includegraphics[width=3cm, height=1cm]{waveform_grey.png}};
            \end{tikzpicture}
        };
    }
    
    \def\initialShiftX{3.5cm} 
    \def\initialShiftY{-2cm} 
    \def\shiftStepX{-0.2cm} 
    \def\shiftStepY{-0.2cm} 
            
    \node (key1) at ([xshift=\initialShiftX+0*\shiftStepX,yshift=\initialShiftY+0*\shiftStepY]outpoint) [anchor=0] {
        \begin{tikzpicture}
            \node [opacity=1*0.2] {\includegraphics[width=1cm, height=1cm]{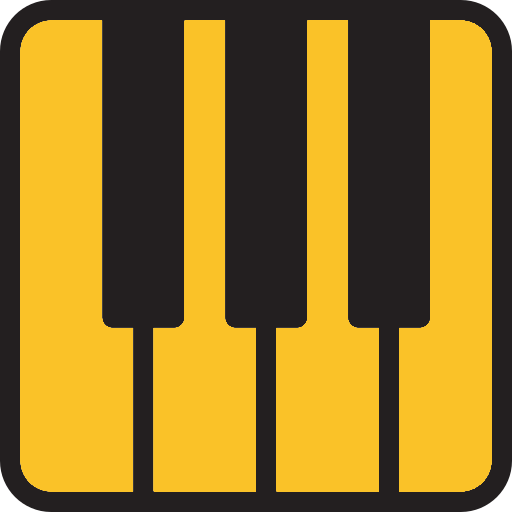}};
        \end{tikzpicture}
    };
    \node (key2) at ([xshift=\initialShiftX+1*\shiftStepX,yshift=\initialShiftY+1*\shiftStepY]outpoint) [anchor=0] {
        \begin{tikzpicture}
            \node [opacity=2*0.2] {\includegraphics[width=1cm, height=1cm]{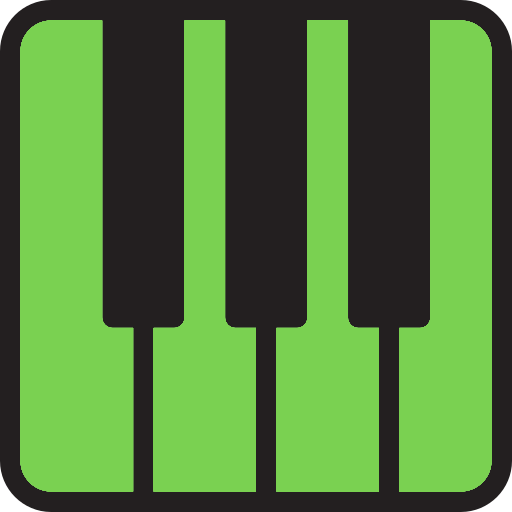}};
        \end{tikzpicture}
    };
    \node (key3) at ([xshift=\initialShiftX+2*\shiftStepX,yshift=\initialShiftY+2*\shiftStepY]outpoint) [anchor=0] {
        \begin{tikzpicture}
            \node [opacity=3*0.2] {\includegraphics[width=1cm, height=1cm]{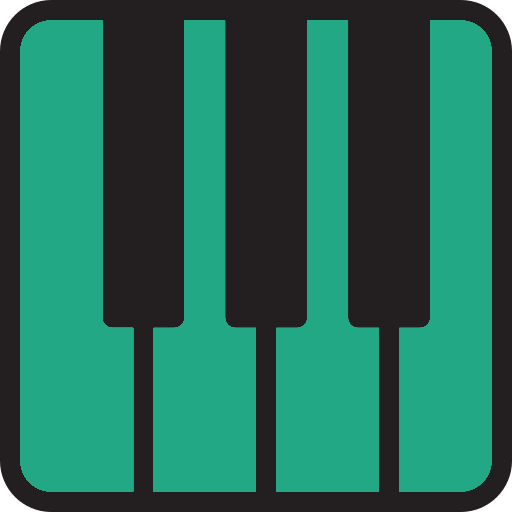}};
        \end{tikzpicture}
    };
    \node (key4) at ([xshift=\initialShiftX+3*\shiftStepX,yshift=\initialShiftY+3*\shiftStepY]outpoint) [anchor=0] {
        \begin{tikzpicture}
            \node [opacity=4*0.2] {\includegraphics[width=1cm, height=1cm]{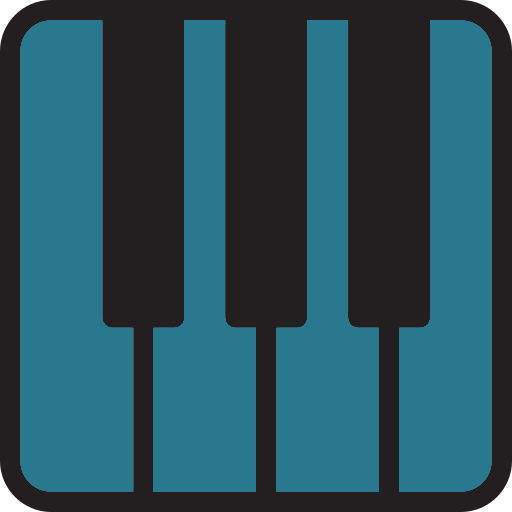}};
        \end{tikzpicture}
    };
    \node (key5) at ([xshift=\initialShiftX+4*\shiftStepX,yshift=\initialShiftY+4*\shiftStepY]outpoint) [anchor=0] {
        \begin{tikzpicture}
            \node [opacity=5*0.2] {\includegraphics[width=1cm, height=1cm]{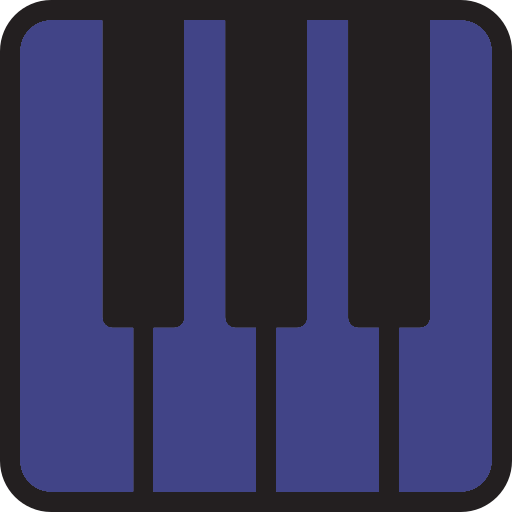}};
        \end{tikzpicture}
    };

    \coordinate (velocity_axis_start) at ($(key5.south) + (0.75cm, 0cm)$);
    \coordinate (velocity_axis_end) at ($(key1.south) + (0.75cm, 0cm)$);

    \coordinate (pitch_axis_start) at ($(key5.south) + (-0.5cm, -0.cm)$);
    \coordinate (pitch_axis_end) at ($(key5.south) + (0.5cm, -0.cm)$);

    \draw [-stealth] (velocity_axis_start) -- (velocity_axis_end) node[midway, below, sloped] {velocity};

    \draw [-stealth] (pitch_axis_start) -- (pitch_axis_end) node[midway, below] {pitch};

    \node[rotate=90, anchor=north] at ($(waveform_output_dac.north) + (-0.1cm, 0cm)$) (waveform_output_dac_text) {output waveform(s)};

    \draw [-stealth] (dac_decoder.east) -- ($(dac_decoder.east -| waveform_output_dac.north)$) node[midway, above] {$\mathbf{\hat{x}}_{k}(p, v)$};

    \draw [-stealth] (dac_decoder.east) -- ($(dac_decoder.east -| waveform_output_dac.north)$) node[midway, below] {$(\mathbf{\hat{X}}_k)$};

    \node (p_linear) [linear, below=5mm of rvq_linear, xshift=-5mm]{\glsentrytext{lut} $\rightarrow$ Linear};
    \node (p) [left=10mm of p_linear]{pitch $p$};
    
    \node (v_linear) [linear, below=1mm of p_linear]{\glsentrytext{lut} $\rightarrow$ Linear};
    \node (v) [left=10mm of v_linear]{velocity $v$};
    
    \node (c_linear) [linear, right=21mm of v_linear]{\glsentrytext{lut} $\rightarrow$ Linear};
    \node (c) [right=10mm of c_linear]{instrument family $f$};
    
    \node (s_linear) [linear, right=21mm of p_linear]{\glsentrytext{lut} $\rightarrow$ Linear};
    \node (s) [right=10mm of s_linear]{source type $s$};

    \node (cat) [linear, right=\dx of rvq_linear] {$\mathrm{cat}(\cdot)$};

    \draw [-stealth, dashed] (dac_encoder) -- (rvq);
    \draw [-stealth, dashed] (rvq) -- (lm) node[midway, below] {$\mathbf{c}$};

    \node[dot] (dot1) at ($(rvq.east)!0.5!(lm.west)$) {};
    \node[dot] (dot2) at ($(lm.east)!0.5!(dac_decoder.west)$) {};

    
   \draw [-stealth] (lm) -- (dac_decoder) node[midway, below] {$\mathbf{\hat{c}}$};


    \draw [-stealth] (rvq_clap) -- (rvq_linear) node[midway, above] {$\bm{\theta}_{\mathrm{\glsentrytext{clap}}}$};
    \draw [-stealth] (rvq_linear) -- (cat);

    \draw [-stealth] (p) -- (p_linear);
    \draw [-stealth] (v) -- (v_linear);
    \draw [-stealth, dashed] (c) -- (c_linear);
    \draw [-stealth, dashed] (s) -- (s_linear);
    
    \draw [-stealth, rounded corners] (p_linear) -| ($(cat.south) + (-0.3cm, 0cm)$) node[near start, above] {$\bm\theta_p$};
    \draw [-stealth, rounded corners] (v_linear) -| ($(cat.south) + (-0.1cm, 0cm)$) node[near start, above] {$\bm\theta_v$};
    \draw [-stealth, rounded corners] (c_linear) -| ($(cat.south) + (0.1cm, 0cm)$) node[near start, above] {$\bm\theta_f$};
    \draw [-stealth, rounded corners] (s_linear) -| ($(cat.south) + (0.3cm, 0cm)$) node[near start, above] {$\bm\theta_s$};
    

    \draw [-stealth, rounded corners] (cat.north) -- (lm.south -| cat.north) node[near end, left] {$\bm{\theta}$} node[near end, right] {cross attention};


    \node[loss, above=5mm of lm] (loss_box) {\(\mathcal{L}_\mathrm{ce}\)};

    \draw [-stealth, dashed, rounded corners] (dot1) |- (loss_box.west);
    \draw [-stealth, dashed, rounded corners] (dot2) |-  (loss_box.east);

  \end{tikzpicture}
  }
  \caption[]{Overview of our proposed system. Dotted lines represent frozen pretrained modules. Dashed lines denote steps exclusive to training. \glsentrytext{clap}'s audio or text head can be used at inference, disregarding source type and instrument family. Training operates on individual samples $\mathbf{x}$, while inference creates a set of samples $\hat{\mathbf{X}}$ from a consistent CLAP prompt and varied pitch/velocity cues to create a full instrument. Different piano keys/colors denote different pitches/velocities.}
  \label{fig:model_overview}
  \vspace{-1.0em}
\end{figure*}

\section{Introduction}\label{sec:introduction}

The exploration of sound synthesis and the development of interfaces to manipulate timbre are fundamental topics in audio research \cite{narita_ganstrument_2023}. With the evolution of sound synthesis in the digital realm, musicians have unprecedented means to manifest their artistic visions. Meanwhile, generative models for images and text have shown disruptive abilities in creating novel samples from learned distributions \cite{barber_muse_2023}. It becomes only natural to consider implications of such technologies when applied to a music production context.

Several generative models for neural audio synthesis have been put forth, including NSynth \cite{engel_neural_2017}, which uses a WaveNet \cite{oord2016wavenet} autoencoder to create samples of pitched instruments, and GANSynth \cite{engel_gansynth_2019}, which models signal phase through an instantaneous frequency representation. Furthermore, \gls{ddsp} \cite{engel_ddsp_2020} and its related works \cite{wu_sawsing_2022} introduce autoencoders with differentiable synthesizers for improved controllability, while a novel approach via a real-time variational autoencoder is presented in \cite{caillon_rave_2021}. Additionally, GANstrument \cite{narita_ganstrument_2023} leverages a feature descriptor obtained through adversarial domain confusion, highlighting the diverse methodologies employed to advance the field of audio synthesis.
These models lack an interface for controlling audio generation via text input. Accordingly, we have witnessed a surge in text-to-audio systems generating convincing audio examples from text prompts \cite{evans_stable_2024}. One family of approaches rely on neural audio codecs \cite{zeghidour_soundstream_2021, kumar_high-fidelity_2023} representing audio as a set of discrete codes whose sequence can be learned using transformer-based language models. While initial approaches targeted speech \cite{borsos_audiolm_2023, wang_valle_2023} and ambient sounds \cite{kreuk_audiogen_2023}, follow-on works adapt methods for text-to-music generating full musical passages from text \cite{agostinelli_musiclm_2023, copet_simple_2023}.

Though compelling, seminal text-to-music works target generation of entire musical arrangements or otherwise lack fine-grained control over their outputs, and might not integrate well into musicians' workflows. Consequently, efforts have been made to adapt these models to sit closer in the creative process. These include StemGen \cite{parker2023stemgen}, predicting instrument track  layers from a given musical context, and VampNet \cite{garcia2023vampnet}, generating musical variations via generative filling. We align with this philosophy, intending to enable new sounds to inspire musical creativity.

In this paper, we introduce the application of neural audio codec language models for the automated creation of sample-based musical instruments using both text and audio prompts as input, building upon our preliminary work in progress in \cite{ml4a2023}. We model a musical instrument as a set of waveforms sampling the instrument's time-domain response across the dimensions of pitch (the fundamental frequency of a note) and velocity (the intensity with which a note is played). Under this paradigm, we move beyond the constraints of any one parametric synthesizer, avoiding expressivity limitations tied to its implementation. As in \cite{narita_ganstrument_2023}, we note that injecting inductive bias into the generative process via \glsentrytext{ddsp} is interesting but complementary to our work, as such methods constrain the manifold that outputs can live on \cite{hayes_ddsp_2024}. Unlike text-to-music systems, which typically generate a single audio example for a given text prompt during inference, prompt-to-instrument systems must generate an ensemble of audio samples from a fixed prompt, which must be pitch-accurate and timbrally consistent with one another to allow for the assembly of a playable instrument. Our contributions are as follows:

• We introduce the \gls{t2i} task, in which waveforms comprising a sample-based musical instrument are generated from a user text prompt.

• We propose neural audio codec language models as solutions for both text- and audio-prompted sample-based instrument generation, expanding on a state-of-the-art generative audio model that is conditioned on a \gls{clap} embedding \cite{wu_large-scale_2023}, as well as pitch across the 88-key range of a standard full-length piano keyboard, velocity, instrument family and source type.

• We develop an objective metric to assess the \gls{tc} of sample-based instruments.

• We propose an adaptation to the average \glsentrytext{clap} score to be suitable for objectively assessing \glsentrytext{t2i}.

• We propose and analyze three \glsentrytext{clap} conditioning schemes through qualitative and quantitative means.

• We demonstrate the compatibility of our approach with both \gls{ar} and non-\glsentrytext{ar} audio transformers like MAGNeT \cite{ziv_magnet_2024}.

The remainder of this paper is organized as follows: Section \ref{sec:method} describes our proposed method, Section \ref{sec:criteria} outlines quantitative metrics for assessing performance, including the ones that we have developed, Section \ref{sec:results} reports our experimental results, and Section \ref{sec:conclusions} draws conclusions.
\section{Proposed method}
\label{sec:method}
Figure~\ref{fig:model_overview} illustrates our proposed method, which is based on MusicGen \cite{copet_simple_2023} as a foundation, consisting of a neural audio codec and a language model to predict acoustic tokens from conditioning signals. We replace EnCodec \cite{defossez_high_2022} used in MusicGen with the \gls{dac} \cite{kumar_high-fidelity_2023}, addressing codebook collapse in previous models while achieving higher audio fidelity. We also introduce a set of new conditioning signals including pitch and velocity, alongside a \glsentrytext{clap} embedding \cite{wu_large-scale_2023}. Our conditioning signals reflect global cues $\bm{\theta}$ for steering generation, which are fused with the language model via cross-attention. Using \glsentrytext{clap} allows instrument samples to be inferred from either audio or text prompts, and we denote their tasks as \gls{s2i} and \gls{t2i}, respectively. The aim of \glsentrytext{s2i} may be considered one of pitch/velocity shifting, whereby the model transforms an audio prompt in ways transcending conventional signal processing. In \glsentrytext{t2i}, text acts as a semantic interface to generate instruments whose timbres may otherwise not exist. To ensure the reproducibility of our findings, we use pretrained sub-networks without modification, training our core language models from random initialization on the standard research dataset NSynth \cite{engel_neural_2017}. We acknowledge that fine-tuning sub-modules within a generative model can improve a composite system, but consider this to be outside the scope of this work.
\subsection{Compressed audio representation}
We use the \glsentrytext{dac} encoder to create an intermediate representation of a monophonic waveform $\mathbf{x}$, resulting in the discrete codes $\mathbf{c}$, while the \glsentrytext{dac} decoder synthesizes an audio waveform $\mathbf{\hat{x}}$ from a predicted code sequence $\mathbf{\hat{c}}$. The \glsentrytext{dac} is trained on a broad spectrum of audio types, so we deem it suitable for generating tonal one-shot instrumental sounds. We model our task at a sample rate of 44.1~kHz, as this would be a minimum requirement for real-world music production use cases. We employ the corresponding pretrained model with fixed weights during training.
\subsection{Language model}
To model the discrete audio tokens of single-shot samples, we consider a smaller, 60M parameter variant of the transformer decoder in \cite{copet_simple_2023}, in order to prevent overfitting, speed up inference, and conceptually demonstrate our approach. The model consists of 12 layers with 16 attention heads per layer and a transformer dimension $d=512$. We consider scaling our models to larger sizes to be out of scope for this work. As in MusicGen \cite{copet_simple_2023}, we predict audio from tokens of the 4 most significant \cite{kumar_high-fidelity_2023} codebooks at each frame (of the 9 supported by \glsentrytext{dac}), selecting tokens from codebooks of size 1024. At inference time, we consider next-token prediction using \glsentrytext{ar} sampling with delayed pattern interleaving \cite{copet_simple_2023}, as well as the iterative decoding scheme proposed in \cite{ziv_magnet_2024} reporting a 7$\times$ inference speed-up. For MAGNeT-style inference, we use 20 decoding steps for the first codebook, and 10 for the remaining codebooks, respectively (compared to 345 steps for the \glsentrytext{ar} scheme). As is customary, we can leverage classifier-free guidance at inference time in both cases \cite{copet_simple_2023, parker2023stemgen}. We expect \glsentrytext{ar} priors to provide higher fidelity, considering the importance of onsets to perception \cite{engel_onsets_2018} for the single-shot samples that we generate: earlier audio token predictions are likely to be perceptually more relevant than later ones.
\subsection{Categorical conditioning} 
We use a categorical conditioning scheme for pitch $p$, velocity $v$, broad instrument family $f$, and source type $s$, that consists of a \gls{lut} and a fully connected layer that maps the dimension of the categorical feature space to the dimension $d$ of the language model. For pitch, we model the $d_p=88$ range of notes spanned by a full-length keyboard, corresponding to \gls{midi} note numbers 21-108, and note this to be a significant expansion relative to the chroma feature used in \cite{copet_simple_2023}. We consider $d_v=5$ velocity layers, according to \glsentrytext{midi} velocities 25, 50, 75, 100, and 127 within our training dataset. The instrument family (i.e. bass, brass, etc.) and source type (i.e., acoustic, electronic, etc.) attributes in our dataset serve as metadata-driven timbral cues that could optionally guide training \cite{vapnik2015learning}, but we do not expect them to be specified at inference. We choose to include them for models trained in this work, subjecting them to dropout with 30\% probability, noting that dropout can generalize their complete inclusion or exclusion.
\subsection{Joint text and audio conditioning}
\label{ssec:clap}
We use the \glsentrytext{clap} model \cite{wu_large-scale_2023}, employing encoders to generate a common fixed-dimensional representation for audio/text pairs of size $d_z=512$. This model was pretrained on musical signals, utilizing a contrastive loss to align respective audio and text embeddings, ultimately enabling the use of either modality as input to our system. The audio encoder $E_\mathrm{a}$ uses HTS-AT \cite{chen_hts-at_2022}, while the text encoder $E_\mathrm{t}$ is based on RoBERTa \cite{liu_roberta_2019}. Given an audio dataset of instrumental samples, this strategy allows for leveraging only the audio head during language model training, without requiring rich text captions in the dataset. We quantize resulting \glsentrytext{clap} embeddings through \gls{rvq} with learned codes \cite{copet_simple_2023}, yielding $\bm{\theta}_{\mathrm{\glsentrytext{clap}}}$.

A distinction between generating music and creating sample-based instruments from prompts is that the inference scenario for instrument generation utilizes a single fixed representation as input for generating a cohesive set of waveforms comprising an instrument. Consequently, we present three \glsentrytext{clap} conditioning schemes specifically to train language models for sample-based instrument creation. These techniques amount to assigning pairs of $\mathbf{z}_\mathrm{\glsentrytext{clap}, a}$ and codes $\mathbf{c}$ as input and target training examples in various ways, where $\mathbf{z}_\mathrm{\glsentrytext{clap}, a}$ is the output of the \glsentrytext{clap} audio encoder $E_\mathrm{a}$. Hence, the target codes and \glsentrytext{clap} embedding within a training example need not be derived from the same waveform, so long as they come from the same instrument. Excluding $\bm\theta_{f}$ and $\bm\theta_{s}$ for clarity, the forward pass observed during the training of a language model $\Theta$ is
\begin{equation}
    \mathbf{\hat{c}}=\Theta(\mathbf{z}_\mathrm{\glsentrytext{clap},a}, \bm\theta_{p}, \bm\theta_{v}),
    \label{eqn:general}
\end{equation}
where $\mathbf{z}_\mathrm{\glsentrytext{clap}, a} = E_\mathrm{a}(\mathbf{x}_k(\rho, \nu))$. Here, $k$, $\rho$, and $\nu$ denote the timbre (i.e. instrument), pitch, and velocity exhibited in an underlying audio example, respectively, which we assume to be readily selectable from our training set. This $\mathbf{x}_k(\rho, \nu)$ is the input to $E_\mathrm{a}$, and need not be identical to $\mathbf{x}_k(p, v)$ which is used to derive the target codes $\mathbf{c}$.
\subsubsection{Baseline \glsentrytext{clap}}
By design, the \glsentrytext{clap} audio encoder $E_\mathrm{a}$ will inevitably yield distinct numerical representations for instrumental samples of the same instrument but varying in pitch or velocity. During training, the following equation applies: 
\begin{equation}
    \mathbf{z}_\mathrm{\glsentrytext{clap}, a} = E_\mathrm{a}(\mathbf{x}_{{k}}(p, v)),
    \label{eqn:vanilla}
\end{equation}
While this suffices for creating a music track from a singular representation, the scenario diverges significantly for sample-based instrument creation. Specifically, pitch and velocity are represented through both the \glsentrytext{clap} representation as well as their respective categorical conditioners, which can reduce the overall effectiveness of the latter. We consider this  adaptation of existing prompt-to-audio methodologies to serve as a baseline in this work, noting its application to this task is still novel.
\subsubsection{Random \glsentrytext{clap}}
In order to disentangle the aforementioned pitch/velocity effect, we consider a randomization technique defined by
\begin{equation}
\mathbf{z}_\mathrm{\glsentrytext{clap}, a} = E_\mathrm{a}(\mathbf{x}_{{k}}(\tilde{{\rho}}, \tilde{{\nu}})),
\label{eqn:random}
\end{equation}
with $\tilde{{\rho}}\!\sim\!\mathcal{U}\{21, ..., 108\}$, and $\tilde{{\nu}}\!\sim\!\mathcal{U}\{25, 50, 75, 100, 127\}$. Random selection with replacement is performed throughout training. This method resembles the nearest neighbor data augmentation in \cite{narita_ganstrument_2023}, where we consider samples to be neighbors if they originate from the same instrument. 
\begin{table}[b]
 \vspace{-1em}
\centering
  \begin{tabular}{lr}
    \toprule
     Instrument families & Note name \\ \midrule
     Bass & C2 \\
     Brass, String, Synth lead & C3 \\
     Guitar, Keyboard, Organ, Reed, Vocal & C4 \\
     Flute, Mallet & C5 \\
     \bottomrule
  \end{tabular}
 \caption{Pitch values used for fixed \glsentrytext{clap} conditioning.}
 \label{tab:fixed}
\end{table}
\subsubsection{Fixed \glsentrytext{clap}}
Lastly, we consider a conditioning scheme where we use a fixed, predefined \glsentrytext{clap} embedding for each instrument as
\begin{equation}
\mathbf{z}_\mathrm{\glsentrytext{clap}, a} = E_\mathrm{a}(\mathbf{x}_{{k}}({\rho}_{0, f}, {\nu}_{0})),
\label{eqn:fixed}
\end{equation}
where ${\rho}_{0, f}$ is defined for each instrument family $f$ (see Table \ref{tab:fixed}) such that fixed representations are sampled within the natural range of each instrument (i.e. we make lower-pitched selections for bass sounds). The categorical velocity ${\nu}_{0}$ is fixed across the training set at velocity 100, conveying an instrument's timbre played with a medium/strong intensity.  If a sample matching a ${\rho}_{0, f}$ and ${\nu}_{0}$ query is not available within an instrument, we opt for its nearest available pitch, followed by its nearest velocity.

Other fixed \glsentrytext{clap} conditioning forms could also have been devised, e.g. using average per-instrument \glsentrytext{clap} embeddings. We opt for our described approach as it ensures that each \glsentrytext{clap} embedding used in model training originates from exactly one audio example.  We assert that this fixed variant most closely aligns training to the scenario at inference. In fact, we posit that both the baseline and random \glsentrytext{clap} approaches are data augmentation alternatives relative to this method, that increase the number of conditioning signal/target code pairs observed during training, while potentially introducing domain mismatches.
\section{Objective Evaluation criteria}
\label{sec:criteria}
We assess models across several objective criteria for \glsentrytext{s2i} and \glsentrytext{t2i}. Alongside the widely used \acrfull{fad} \cite{kilgour2018fr} score, we introduce a novel metric to evaluate the \glsentrytext{tc} of generated sample-based instruments. We also propose an adaptation of the average \glsentrytext{clap} score to fairly evaluate text correspondence for \glsentrytext{t2i}. Unless otherwise specified, we base instrument generation-specific metrics on the assumption that they are represented by $N_k = d_{p}d_{v} = 440$ audio samples. In practice, care is taken to properly aggregate/mask instrument statistics based on which samples are present.
\subsection{\glsentrytext{fad} score}
The \glsentrytext{fad} score allows a common framework for evaluating generative audio models using almost any audio feature descriptor \cite{kilgour2018fr}. We utilize a \glsentrytext{fad} metric formulated using VGGish, as in related works \cite{agostinelli_musiclm_2023, parker2023stemgen}. We also report \glsentrytext{fad} scores using \glsentrytext{clap} (audio) embeddings, since they form a pivotal component to our system, allow analysis for higher-sample rate audio (48~kHz), and have been shown to have increased correlation to perception relative to VGGish \cite{gui2023adapting}. The \glsentrytext{fad} score is generically defined as
\begin{align}
\label{eqn:fad}
\mathrm{FAD}\left(\mathbf{Z}_1, \mathbf{Z}_2\right) &= \lVert \mathbf{\mu}_1 - \mathbf{\mu}_2 \rVert_2^2 \nonumber \\
    &\quad + Tr\left(\mathbf{A}_1 + \mathbf{A}_2 + \left(\mathbf{A}_1 \mathbf{A}_2\right)^{\frac{1}{2}}\right),
\end{align}
where $\mathbf{Z}_i \in \mathbb{R}^{d_z \times TN}$ is a collection of $T$ $d_z$-dimensional embeddings extracted by a given audio descriptor, across $N$ samples from a population $i \in [1, 2]$. Considering the 4-second long audio segments generated in this work and the strides of various models, $T=4$ and $1$ when using VGGish and \glsentrytext{clap}, respectively. We reserve subscripts $1$ and $2$ to denote ground truth/test populations, respectively. Accordingly, each $\mathbf{Z}_i$ has mean $\mathbf{\mu}_i \in \mathbb{R}^{d_z}$ and covariance $\mathbf{A}_i \propto \mathbf{Z}_{i}\mathbf{Z}_{i}^\top \in \mathbb{R}^{d_z \times d_z}$. The first and second terms in Equation \ref{eqn:fad} quantify mean correspondence and similarities in the spread between distributions, respectively. The \glsentrytext{fad} score possesses a property allowing unpaired populations to be compared, which we use as a criterion to assess "in-the-wild" \glsentrytext{t2i} in lieu of ground truth audio.
\begin{figure*}[hbt]
\centering
  \begin{minipage}{.235\textwidth}
	\centering
  \centerline{\includegraphics[width=0.95\columnwidth]{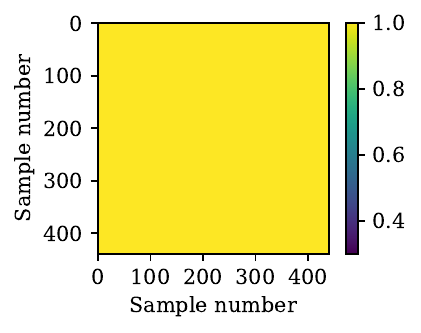}}
	\vspace{0.0em}(a)
  \end{minipage}
  \begin{minipage}{.235\textwidth}
	\centering
  \centerline{\includegraphics[width=0.95\columnwidth]{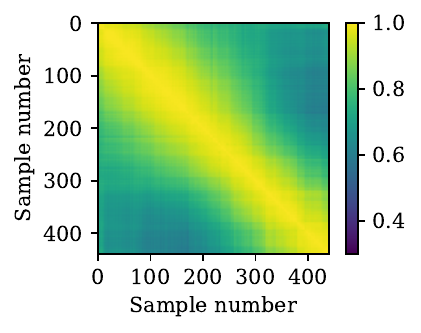}}
	\vspace{0.0em}(b)
  \end{minipage}
  \begin{minipage}{.235\textwidth}
	\centering
  \centerline{\includegraphics[width=0.95\columnwidth]{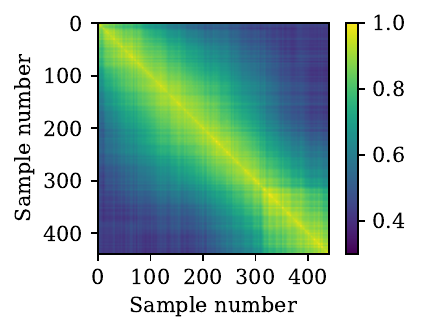}}
	\vspace{0.0em}(c)
  \end{minipage}
  \begin{minipage}{.235\textwidth}
	\centering
  \centerline{\includegraphics[width=0.8\columnwidth]{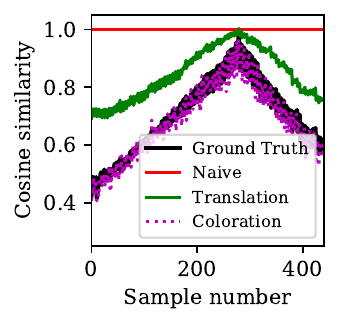}}
	\vspace{0.0em}(d)
  \end{minipage}
\caption{Covariance matrices for the text prompt $\mathbf{t}_k=$ \texttt{aggressive synth lead}, computed using (a) naive replication, (b) translation, (c) coloration (matching the ground truth covariance $\mathbf{A}_{11,*}$ learned over the 53 instruments reflected in the NSynth validation/test sets), (d) cosine similarities relative to estimated $\hat\rho_k$/$\hat\nu_k$, corresponding to note E5/velocity 100.}
\vspace{-1.0em}
\label{fig:cov}
\end{figure*}
\subsection{TC score}
\label{ssec:tc}
Our system should generate timbrally consistent samples in order for them to triggered harmoniously as a sample-based instrument, and we aim to characterize this quantitatively. An apt definition for \glsentrytext{tc} may seem ill-posed, since we want instrument samples to be fundamentally consistent with one another, but also expect them to exhibit some timbral variations as functions of pitch/velocity. This is particularly sought-after in high-quality virtual instruments, motivating the modeling approach in \cite{engel_ddsp_2020}. To contend with these potentially conflicting aspirations, we learn statistics from existing sample-based instruments serving as prototypes for realistic \glsentrytext{tc}, and build metrics around them. We use \glsentrytext{clap} embeddings as a basis to create an elegant embodiment in this work. To do so, we forego the mean subtraction step standard to covariance matrix computations, noting that samples are practically close to zero-mean in this respect. Hereafter, we use the terms covariance, affinity, and cosine similarity interchangeably.

We define per-instrument covariance matrices as
\begin{equation}
\label{eqn:affinity_inst}
    \mathbf{A}_{ij,k} = \frac{1}{N_k} \mathbf{Z}_{i,k}^\top \mathbf{Z}_{j,k},
\end{equation}
where $\mathbf{A}_{ij,k} \in \mathbb{R}^{N_k \times N_k}$ is the affinity between embeddings $\mathbf{Z}_{i,k}$ and $\mathbf{Z}_{j,k} \in \mathbb{R}^{d_z \times N_k}$ representing the subset of \glsentrytext{clap} embeddings of the $k$th instrument within each population. Here, we compute statistics emphasizing variations across samples instead of feature dimensions. Referring to Equation \ref{eqn:fad}, the $L_2$-normalized quality \glsentrytext{clap} embeddings will ensure us that $Tr\left(\mathbf{A}_{ii,k}\right)=1$ $\forall~$ $i \in [1, 2]$ and $k \in [1, \dots, K]$.  Accordingly, we can define
\begin{equation}
\label{eqn:tc}
\mathrm{\glsentrytext{tc}}_{\mathrm{\glsentrytext{clap}}}\left(\mathbf{Z}_1, \mathbf{Z}_2\right) = \frac{1}{K}\sum_{k}^{K}Tr\left(\left(\mathbf{A}_{11,k}\mathbf{A}_{22,k}\right)^{\frac{1}{2}}\right),
\end{equation}
which is bounded in [0, 1] and aggregates the similarity in covariations across instruments within each population. Instead of using $\mathbf{A}_{11,k}$ for making comparisons between populations on a per-instrument basis, we consider $\mathbf{A}_{11,*} = \frac{1}{K}\sum_{k}^{K}\mathbf{A}_{11,k}$, averaging per-instrument affinity matrices across a ground truth evaluation set. This provides richer statistics for improved stability, and a unified method to assess \glsentrytext{tc} for \glsentrytext{s2i} and \glsentrytext{t2i}. The \glsentrytext{tc} score is then
\begin{equation}
\label{eqn:tc_learned}    \mathrm{\glsentrytext{tc}}_{\mathrm{\glsentrytext{clap}*}}\left(\mathbf{Z}_1, \mathbf{Z}_2\right) = \frac{1}{K}\sum_{k}^{K}Tr\left(\left(\mathbf{A}_{11,*}\mathbf{A}_{22,k}\right)^{\frac{1}{2}}\right).
\end{equation}

We compute $\mathbf{A}_{11,*}$ using all of the samples from the NSynth validation and test sets that are within our desired 88-key pitch range, reflecting a total of 53 instruments.  The resulting covariance matrix is illustrated in Figure \ref{fig:cov}c, in which samples are ordered primarily by pitch and secondarily by velocity.  Note how $\mathbf{A}_{11,*}$ deviates from "ideal \glsentrytext{tc}," whereby all embeddings would be correlated with unity similarity (see Figure \ref{fig:cov}a).  Moreover, a $5\times 5$ texture emerges in $\mathbf{A}_{11,*}$, indicative of variations in cosine similarity amongst samples of the same pitch but differing velocities.  Lastly, one may question the suitability of \glsentrytext{clap} as a feature descriptor within this context, given its variability concerning pitch/velocity discussed in Section~\ref{ssec:clap}. Its improved correlation to perception aside \cite{gui2023adapting}, we assert that learning statistics over data effectively embeds potential measurement deficiencies that effectively neutralizes when we compare new population statistics against it.
\subsection{Average \glsentrytext{clap} score}
\subsubsection{Sample-to-instrument (\glsentrytext{s2i})}
Given $N = \sum_{k}^{K}N_k$ and a cross-population covariance $\mathbf{A}_{ij} = \frac{1}{N}\mathbf{Z}_{i}^\top \mathbf{Z}_{j} \in \mathbb{R}^{N \times N}$, the average \glsentrytext{clap} score computed on a per-sample basis can be expressed concisely as 
\begin{equation}
\label{eqn:avg_clap_samp}
    s_{\mathrm{\glsentrytext{clap}}}\left(\mathbf{Z}_1, \mathbf{Z}_2\right) = Tr\left(\mathbf{A}_{12}\right) = \frac{1}{N}\sum_{k}^{K}N_kTr\left(\mathbf{A}_{12,k}\right).
\end{equation}
It can also be computed on a per-instrument basis by
\begin{equation}
\label{eqn:avg_clap_inst}
    s_{\mathrm{\glsentrytext{clap}}*}\left(\mathbf{Z}_1, \mathbf{Z}_2\right) = \frac{1}{K}\sum_{k}^{K}Tr\left(\mathbf{A}_{12,k}\right).
\end{equation}
We opt for this version in our work, noting that the two measures are equivalent when $N_1 = N_2 = \dots = N_K$.  
\begin{table*}[t]
\centering
  \begin{tabular}{lccccc}
    \toprule
     Model & Inference &\glsentrytext{fad}$_{VGGish}\downarrow$ & \glsentrytext{fad}$_{\glsentrytext{clap}}\downarrow$ & $s_{\glsentrytext{clap}*}\uparrow$ & \glsentrytext{tc}$_{\glsentrytext{clap}*}\uparrow$ \\ \midrule
     Baseline \glsentrytext{clap} & \glsentrytext{ar} & 1.781 & 0.214 & 0.626 & 0.937 \\
     Random \glsentrytext{clap} & \glsentrytext{ar} & \textbf{1.558} & \textbf{0.196} & \textbf{0.656} & 0.929 \\
     Fixed \glsentrytext{clap} & \glsentrytext{ar} & 1.951 & 0.225 & 0.637 & \textbf{0.951} \\ \midrule
    Baseline \glsentrytext{clap} & MAGNeT &  1.974 & 0.263 & 0.561 & 0.931 \\ \bottomrule
  \end{tabular}
 \caption{Objective \glsentrytext{s2i} evaluation over the NSynth test set.}
 \label{tab:s2i}
 \vspace{-0.5em}
\end{table*}
\begin{table*}[ht]
\centering
\begin{tabular}{lccc||cccc}
\toprule
Model & \glsentrytext{fad}$_{VGGish}\downarrow$ & \glsentrytext{fad}$_{\glsentrytext{clap}}\downarrow$ & \glsentrytext{tc}$_{\glsentrytext{clap}*}\uparrow$ & Naive & Translation & Coloration \\
\midrule
Baseline \glsentrytext{clap} & 3.060 & 0.402 & 0.908 & \textbf{0.225} & \textbf{0.239} & 0.359 \\
Random \glsentrytext{clap} & \textbf{2.416} & \textbf{0.315} & 0.883 & 0.168 & 0.224 & \textbf{0.361} \\
Fixed \glsentrytext{clap} & 3.668 & 0.427 & \textbf{0.932} & 0.171 & 0.204 & 0.333 \\
\bottomrule
\end{tabular}
\caption{Objective \glsentrytext{t2i} evaluation over a curated set of text prompts (left), and using $s_{\mathrm{\glsentrytext{clap}}*}\uparrow$ comparing naive application of \glsentrytext{clap} text embeddings against the proposed translation and coloration methods for synthesizing $\mathbf{Z}_{1,k}$ (right).}
\vspace{-1.0em}
\label{tab:combined}
\end{table*}
\subsubsection{Text-to-instrument (\glsentrytext{t2i})}
\label{ssec:t2i_avg_clap}
The average \glsentrytext{clap} score $s_{\mathrm{\glsentrytext{clap}}*}$ is suitable for cases with a one-to-one match between ground truth prompts and their corresponding audio examples. However, it can deteriorate for \glsentrytext{t2i}, where a single \glsentrytext{clap} text embedding must be related to an ensemble of \glsentrytext{clap} audio embeddings $\mathbf{Z}_{2,k}$. A naive adaptation involves comparing each audio embedding within the generated instrument to the same target text embedding.  This amounts to creating $\mathbf{Z}_{1,k}$ by replicating the \glsentrytext{clap} text embedding $N_k$ times (whose resulting covariance is the "ideal \glsentrytext{tc}" one in Figure \ref{fig:cov}a), and using it as input to Equation \ref{eqn:avg_clap_inst}.  Hence, we set out to \emph{synthesize} a realistic ensemble of \glsentrytext{clap} embeddings $\mathbf{Z}_{1,k}$ from a single \glsentrytext{clap} text embedding $\mathbf{z}_\mathrm{CLAP, t} = E_\mathrm{t}(\mathbf{t}_k)$, derived from the $k$th text prompt $\mathbf{t}_k$.  Again, we accomplish this by leveraging statistics from available instrument data.

We construct $\mathbf{M}_{1,*} \in \mathbb{R}^{d_z \times d_pd_v}$ as the mean \glsentrytext{clap} audio embeddings at each pitch/velocity pair across all instruments in our evaluation data, re-normalizing them upon averaging.  We posit that a text prompt implies a specific pitch/velocity (e.g., "softly plucked upright bass" suggests a low pitch/velocity). To estimate the corresponding pitch $\hat\rho_k$ and velocity $\hat\nu_k$ for a given prompt, and to identify its closest template $\mathbf{\hat\mu}_{1,k}$, we use $\mathbf{M}_{1,*}$ as a template matching-based classifier onto $\mathbf{z}_\mathrm{CLAP, t}$.  Accordingly, we can define 
\begin{equation}
\label{eqn:trans}
\mathbf{M}_{1,k}=\mathbf{M}_{1,*}+(\mathbf{\hat\mu}_{1,k}-z_\mathrm{CLAP, t})
\end{equation}
such that $\mathbf{M}_{1,k}$ is aligned to $\mathbf{z}_\mathrm{CLAP, t}$ at $\hat\rho_k$/$\hat\nu_k$. Re-normalizing, we have $\mathbf{Z}_{1,k}=\mathbf{M}_{1,k}/||\mathbf{M}_{1,k}||$. Figure \ref{fig:cov}b illustrates a covariance matrix derived from this approach for a given text prompt.  This \emph{translation} method improves upon naive replication, but contains higher cross-correlations than in $\mathbf{A}_{11,*}$. Finally, we derive a \emph{coloration} transformation $\mathbf{Z}_{1,k}\leftarrow Y(\mathbf{Z}_{1,k}, \mathbf{A}_{11,*})$ through standard Eigendecomposition techniques, resulting in a $\mathbf{Z}_{1,k}$ with covariance $\mathbf{A}_{11,*}$, as in Figure \ref{fig:cov}c. 
\section{Experimental results}
\label{sec:results}
We train models on the NSynth dataset \cite{engel_neural_2017}, pruning it according to our specified 88-key pitch range. We resample the 16~kHz dataset to 44.1~kHz, viewing it as a proxy in lieu of an equally comprehensive full-band alternative. Models are trained to minimize the cross-entropy $\mathcal{L}_\mathrm{ce}$ between predicted codes $\mathbf{\hat{c}}$ and ground truth $\mathbf{c}$, over 1M training steps with AdamW optimizer, a batch size of 48, and a cosine-annealed schedule as in \cite{copet_simple_2023} with an initial learning rate of $10^{-3}$.  We primarily analyze the impact of the proposed \glsentrytext{clap} conditioning training variants with \glsentrytext{ar} inference.  Additionally, we train a baseline \glsentrytext{clap} model with MAGNeT-style iterative decoding to compare its relative performance. To promote consistency in generated samples used for evaluation, we fix the random seed of our categorical samplers, ensuring that generations undergo the same random sampling trajectory. We refer readers to our supplementary materials available at {\url{https://gen-inst.netlify.app/}.

We evaluate and analyze the models through several means. We liken \glsentrytext{s2i} to a reconstruction of the NSynth test set \cite{narita_ganstrument_2023} adapted to our inference condition, as a user can provide a sample at any pitch/velocity available to them and models must render its timbre over all pitch/velocity queries. We simulate this by randomly selecting a single query \glsentrytext{clap} audio embedding for each instrument, using it to generate all other samples within the instrument. For \glsentrytext{t2i}, we curate 25 text prompts of varying complexity, generating instruments accordingly.

\subsection{Objective evaluation}
\label{ssec:obj_eval}
We analyze generations across \glsentrytext{s2i} and \glsentrytext{t2i}, using \glsentrytext{fad} (for overall expressivity and fidelity), $s_{\mathrm{\glsentrytext{clap}}*}$ (for prompt correspondence), and $\mathrm{\glsentrytext{tc}}_{\mathrm{\glsentrytext{clap}*}}$ (for \glsentrytext{tc}) to evaluate models quantitatively. To compute \glsentrytext{fad} scores for \glsentrytext{t2i}, we relate generated instruments to the NSynth test set in the absence of the ground truth audio.  Lastly, we compare the different $s_{\mathrm{\glsentrytext{clap}}*}$ versions for \glsentrytext{t2i} introduced in Section \ref{ssec:t2i_avg_clap}.

Quantitative results for \glsentrytext{s2i} and \glsentrytext{t2i} are summarized in Tables \ref{tab:s2i} and \ref{tab:combined}, respectively.  For \glsentrytext{s2i}, the random \glsentrytext{clap} variant outperforms other models in terms of \glsentrytext{fad} and $s_{\mathrm{\glsentrytext{clap}}*}$ at the expense of reduced \glsentrytext{tc}. The converse is true for the fixed \glsentrytext{clap} variant, which outperforms in \glsentrytext{tc}.  While we do not prescribe which factor is most crucial to overall instrument quality, we do assert that \glsentrytext{tc} is an important element for overall playability. The baseline \glsentrytext{clap} approach slots itself in the middle with regards to all criteria.  Its MAGNeT variant exhibits degraded performance, but generates samples with 7$\times$ fewer inference steps.  These findings are largely mirrored in the \glsentrytext{t2i} case. Interestingly, the baseline \glsentrytext{clap} variant seemingly outperforms models in terms of $s_{\mathrm{\glsentrytext{clap}}*}$ using a naively adapted measure.  The translation method increases scores across all models.  Lastly, we see that the random \glsentrytext{clap} model (marginally) outperforms other variants when using the coloration method, in line with \glsentrytext{s2i}. Note that this version of the measure significantly bolsters $s_{\mathrm{\glsentrytext{clap}}*}$ across all models relative to naive replication and translation, so we argue that it is best-suited for computing \glsentrytext{t2i} $s_{\mathrm{\glsentrytext{clap}}*}$.

\subsection{Subjective evaluation}
\label{ssec:sub_eval}
We used the \gls{mushra} and \gls{mos} methods \cite{camp2023mos} to evaluate model variants subjectively. The \glsentrytext{mushra} test was catered to \glsentrytext{s2i}, and involved participants rating the quality of individual samples generated by different models against a hidden reference (i.e. a ground truth sample) and an anchor (i.e. a sample generated by a randomly initialized model). We performed a 1-5 Likert scale \glsentrytext{mos} test for \glsentrytext{t2i} scenarios, where participants evaluated the audio outputs generated from text prompts based on overall playability and \glsentrytext{tc}. Our accompanying website demonstrates the nature of trials used in our evaluation. 

In total, 62 participants took part in our two-phase evaluation, with results summarized in Table~\ref{tab:listening}. Note that most participants possess expert listening skills and have been involved in virtual instrument creation for several years, contributing to slightly lower absolute results than anticipated.  Listening test results were consistent with our objective evaluation, confirming the two assertions of our work: (1) random \glsentrytext{clap} improves expressivity over baseline \glsentrytext{clap} by virtue of its data augmentation, and (2) fixed \glsentrytext{clap} improves \glsentrytext{tc} over baseline \glsentrytext{clap} because its training more closely resembles the inference condition.

\begin{table}[h]

\centering
\vspace{-0.0em}
  \begin{tabular}{lcc}
    \toprule
     Model & MUSHRA & MOS \\ \midrule
     Baseline \glsentrytext{clap} & 56.08 & 2.290 \\
     Random \glsentrytext{clap} & \textbf{63.35} & 2.661 \\
     Fixed \glsentrytext{clap} & 57.96 & \textbf{2.820} \\ \midrule
     Ground truth & \underline{\textbf{98.45}} & -- \\ 
     Anchor & 0.442 & -- \\ \bottomrule
  \end{tabular}
 \caption{Summary of our subjective listening tests.}
 \vspace{-2em}
 \label{tab:listening}
\end{table}

\section{Conclusions}
\label{sec:conclusions}
In this work, we proposed methods for generating sample-based musical instruments from text or audio prompts using neural audio codec language models. We considered different \glsentrytext{clap} conditioning variants based on the unique challenge of our task, whereby a set of samples that are timbrally consistent must be generated from a single prompt. We proposed metrics to assess sample-based instruments through various means. Extensive evaluations showcased the effectiveness of our methods, underscoring a compromise between expressivity and \glsentrytext{tc}. Future work will enable deeper control for sample generation, where adapters could be used to augment a base model \cite{barber_styledrop_2023}. We would also like to improve system fidelity, scaling models to larger sizes with fine-tuned modules \cite{evans_stable_2024}.

\clearpage
\section{Ethics Statement}

We have intentionally pursued this task as a topic for scientific research as an alternative to more conventional prompt-to-media systems. The spirit of this work is specifically to expand sound synthesis possibilities for music creators in order to realize their artistic visions.  Moreover, we feel that our resulting system and its intents pose far less risk to personal attack/misrepresentation as well as the livelihood of creatives, and is less susceptible to incrimination/impersonation attempts relative to the forms of generative models that have caused increased levels of concern within the general population \cite{genai2023}.

Beyond our primary ethical concerns, we also recognize the environmental implications of our computational practices. Our experiments were carried out using Amazon Web Services in the \textit{us-gov-east-1} region, with a carbon efficiency of 0.57~kgCO$_2$eq per kilowatt-hour. One training of our model entailed approximately 96 hours of computation on Intel Xeon E5-2686 v4 (Broadwell) hardware using a single V100 GPU, culminating in an estimated total emission of 7.93~kgCO$_2$eq. This estimation was facilitated by the Machine Learning Impact calculator \cite{lacoste2019quantifying}. In acknowledging our environmental impact, we underscore the importance of integrating sustainability considerations into the research process, reflecting on the imperative to balance innovation with ecological responsibility.

\bibliography{ISMIRtemplate}

\end{document}